\newcommand{\be}{\begin{equation}}
\newcommand{\ee}{\end{equation}}
\newcommand{\bee}{\begin{eqnarray}}
\newcommand{\eee}{\end{eqnarray}}
\newcommand{\Z}{{\cal Z}}
\newcommand{\G}{{\cal G}}
\newcommand{\D}{{\cal D}}
\newcommand{\Intg}{\int_{-\infty}^\infty}

\documentstyle[twocolumn,prl,aps]{revtex} % for PRB use prb insted of prl
\input epsf
\begin{document}
\draft                         %  -- makes ``pacs'' numbers print
% \preprint{BGU-PH-97/}        % -- will appear in preprint mode.
\wideabs{ %  - to produce wide abstract like in PRB
\title{Exact non-equilibrium current from the partition function  \\
for impurity transport problems}
\author{Sergei Skorik}
\address{Department of Physics, Ben-Gurion University, Beer-Sheva 84105,
Israel \\
and Department of Physics, Weizmann Institute for Science, Rehovot 76100,
Israel
}
\date{30 July 1997}
\maketitle
\begin{abstract}
We study the partition functions  of quantum impurity problems in
the  domain of complex applied bias for its relation  
to the non-equilibrium current suggested by Fendley, Lesage and Saleur
(cond-mat/9510055).
The problem is reformulated as a certain
generalization of the linear response theory that accomodates an additional
complex  variable. 
It is shown that the mentioned relation holds in a rather
generic case in the linear response limit, or under certain condition
out of equilibrium. This condition is trivially satisfied
by the quadratic Hamiltonians and is rather restrictive for the interacting
models. An example is given when the condition is violated.

\end{abstract}
}  % - to close wideabs bracket
%\narrowtext
% \widetext

\section{Introduction}    %\protect\\ }
\label{sec:level1}

Strongly correlated quantum systems out of equilibrium is one
of the lesser understood domains in theoretical physics, mainly
because such systems
reveal  much more complex behavior  compared to the case of 
equilibrium.  
Correspondinly, there are not that many non-trivial examples known 
today that are solved exactly out of equilibrium, and  extension of 
powerful
field-theoretical methods to non-equilibrium situations is desirable.

Situation is somewhat simpler in the stationary non-equilibrium regime.
In a remarkable work, Ref.\ \cite{FLS}, the authors put forward
a novel idea of how to calculate the current for integrable
systems out of equilibrium. In particular, they obtained,
based on the standard Bethe ansatz technique,
a non-perturbative expression for the backscattering 
current through a single impurity
in the quantum Hall bar with $\nu=\case{1}{3}$. Their result, cast into
appropriate form \cite{FLesS}, reads:
\be
I_B={e\over h} i\pi T\nu\lambda_{bs}{d\over d\lambda_{bs}}
\log\left[{\Z_{imp}(i\mu_N\to\mu)
\over\Z_{imp}(-i\mu_N\to\mu)}\right], \label{Sal}
\ee
where $T$ stands for temperature, $\nu=\case{1}{3}$ is a filling fraction,
$\lambda_{bs}$ is an effective tunneling strength of the impurity, $\Z_{imp}$
is {\it impurity's partition function} defined as the ratio of the 
equilibrium partition
function of the full system to that of the ``pure'' system without
impurity (the leads), and  analytic continuation from the discrete 
imaginary to real
chemical potential of the edges $\mu$ is assumed. 
It is worth emphasizing that Eq.\ (\ref{Sal}) was meant to have
 a truly non-equilibrium
nature, unlike similar expressions for persistent currents \cite{BY,Bl},
where the current $I=c\partial F/\partial\Phi$ is a purely equilibrium
phenomenon, a consequence of the incident flux $\Phi$ on the 
structure of energy levels. 

Relation (\ref{Sal}) was obtained 
through several intermediate steps, rather than derived directly.
(a) First, the ``fusion'' of Bethe ansatz with the Bolzmann rate equation
was implemented as follows \cite{FLS}.
The quantum Hall bar with impurity, described by the Luttinger
theory with point-like backscattering term $\lambda_{bs}
\delta(x)(\Psi_L^+\Psi_R+\Psi_R^+\Psi_L)$ 
was reformulated in terms of the boundary sine-Gordon model 
\be
H_{BSG}={1\over 2}\int_0^\infty\!\!\!\! dx[\Pi^2(x)+(\partial_x\Phi)^2]+
\lambda_{bs}\cos{\sqrt{8\pi\nu}\over 2}\Phi(0), \label{BSG}
\ee
which is exactly solvable in equilibrium 
and can be diagonalized by the Bethe ansatz.
Transport of Laughlin $(e/3)$ quasiparticles through impurity 
\cite{Reznik} maps onto  
a scattering of sine-Gordon quasiparticles (kinks, antikinks and breathers)
off the boundary. All the interaction in (\ref{BSG}) is now at the boundary,
which behaves like a non-elastic scatterer: kink can be bounced as 
an antikink and vice versa, which changes the charge of the system. 
However, it is
a bare Hamiltonian where the bulk interactions are absent. As a result
of diagonalization,
sine-Gordon quasiparticles interact also in the bulk by a pairwise
point-like interaction that adds merely a phase-shift with momentum
of each quasiparticle preserved -- a consequence of the peculiar 
conservation
laws of (\ref{BSG}), and the distribution function of the gas of
quasiparticles differs from the usual Fermi one.  
Under these circumstances, the current was calculated 
on the basis of the standard probability arguments which are
known to lead to the (classical) Bolzmann rate equation \cite{FLS}:
\be 
I_B={e\over h}\int_0^\infty \!\! dp[n_+(p,\mu,T)-n_-(p,\mu,T)]
|S_+^-(p,\lambda_{bs})|^2.
\label{Bolz}
\ee
This expression was referred in literature to as an  exact result
because of the exact quantum nature of all the entries
of expression (\ref{Bolz}) used in \cite{FLS}. 
In particular, 
the densities of states $n_{\pm}$
of quasiparticles in the presence of the external bias $\mu\int_0^\infty\partial_x
\Phi$
were found from the thermodynamic Bethe ansatz
(TBA) \cite{FLS,AlZamo}, whereas  
for the transition probability $W=|\langle +|-\rangle|^2$
out of equilibrium
it was taken 
the kink-antikink equilibrium scattering matrix
element $S^-_+$ \cite{Com}, obtained in \cite{Zamo}.

(b) Second, the exact impurity partition function of (\ref{BSG}) was
obtained either as a Coulomb gas expansion, or based on the relation
to the Kondo model in a magnetic field and using TBA for the latter
\cite{FLesS}. Eventually, formula (\ref{Sal}) was checked by comparing
series expansions of both sides to a high order \cite{FLesS,conj}.

Existence of valid
generic relations like Eq.\ (\ref{Sal})
would be rather helpful. Indeed, a partition function is much easier
to find and can be tackled by a variety of techniques: Feynmann
integration in imaginary time, Monte Carlo simulations, Bethe ansatz etc. 
Then the proper analytic continuation in $\mu$ must be done. Speaking
loosely, it is in some sense analogous to the analytic continuation
from Matsubara frequencies to the real axis in the standard linear response
theory. Out of equilibrium, the real time retarded Green function
has no obvious thermodynamic interpretation on the imaginary
 frequency axis; equations of the kind (\ref{Sal}) offer such an 
interpretation. Eq.\  (\ref{Sal}) allows to cook up the non-equilibrium
retarded Green function in the complex frequency plane from the Matrsubara
Green functions in a very simple way. The important question is, however,
how general  formula (\ref{Sal}) is.

In this paper we undertake further studies of the relations of the form
of Eq.\ (\ref{Sal}) without appealing to the knowledge of Bethe ansatz, 
but rather on the basis of standard Green functions technique.
We focuse on the transport properties of one-dimensional electrons passing
through an impurity or an interacting region, the quantity of interest
being an I-V characteristic.
Experimentally, such systems are realized, e.\ g.\ , as quantum dots,
 quenched nanoconstrictions, quantum wires with an impurity or the edges
of quantum Hall liquid. On the theoretical level, we start with
a hypothetical model of two reservoirs, $L$ and $R$, of free
1D electrons with a linearized spectrum, kept at different constant chemical
 potentials $\mu_L$ and $\mu_R$, and described by the Hamiltonian
\be
H_0=H_L+H_R=
\sum_{m=L,R}i\int_{-\infty}^{+\infty}\! dx\psi^+_m\partial_x\psi_m,
\label{Hfree}
\ee
where we have chosen left-moving branch for both $L$ and $R$ species
and set $v_F=1$ \cite{rem1}
(similar starting point of view was taken in 
\cite{FLS}, the  reservoirs being the edge channels or kinks-antikinks
in the sine-Gordon formulation). 
 Each of the reservoirs 
(referred to as {\it leads} below)
is intially in a well-defined equilibrium state.
 Then, one allows for a current
to flow from $L$ to $R$ by adiabatically switching on some coupling term. Since
the amount of electrons in reservoirs is infinite in the infinite
bandwidth limit, the reservoirs never
equilibrate and the system achieves some non-equilibrium steady state
with a constant flux in the energy space. Equation for an exact quantum
current can be derived rigorously in the spirit of \cite{Nozier}. 

To be specific, we consider  three types of coupling terms:
point-contact, resonant-level and Kondo.  
Interactions can be introduced in the  free (quadratic) models (a) 
by moving
away from the Toulouse point in the Kondo model; (b) by taking
leads to be the Luttinger liquid instead of free electrons 
in the point-contact model and
(c) by considering
a larger space of interacting states $\{d_n\}$ instead of one state
in the resonant-level model. We study in detail only the latter case here.   
We argue that: a) equation (\ref{Sal}) holds for free systems (quadratic
couplings), and
b) Eq.\ (\ref{Sal}) holds for the interacting systems under 
certain condition for the self-energy, or as a generic formula
in the linear response limit.
The results can be easily  generilized to the case of transport
through an interacting region if one interprets $\{d_n\}$ as a set of eigenstates
in  the isolated region labeled by the quantum number $n$ with the hopping
strength depending  on $n$. 

\section{Partition function \lowercase{vs.} non-equilibrium current}
\label{sec:inter}

Consider the following Hamiltonian, which describes the leads ($\psi$),
the impurity ($d$) located at $x=0$, and possible interactions ($H_{int}$):
\bee
H=H_0&+&\lambda\!\!\!\!
\sum_{m=L,R}(\psi^+_m(0)d+d^+\psi_m(0)) \nonumber \\
&+& \epsilon_0d^+d + H_{int}(\{d\}),
\label{Hamilton}
\eee
 Operator $d^+$ creates an electron on the impurity site at the energy $\epsilon_0$. 
For this model, an analogue of relation (\ref{Sal}) to be shown reads:
\be
J={e\over h}i\pi T\Gamma{\partial\over\partial\Gamma}\log{
\Z_{imp}^+\over\Z_{imp}^-},
\label{eq:hui}
\ee
where $\Gamma\equiv\lambda^2$ \cite{rem22}.

Define the impurity partition function as
\be
\Z_{imp}(i\mu_{L},i\mu_R)={\text{Tr}\, e^{-(H+i\mu_LN_L+i\mu_RN_R)/T}\over
\text{Tr}\, e^{-(H_0+i\mu_LN_L+i\mu_RN_R)/T}   },
\label{DefPF}
\ee
where $i\mu_{L(R)}$ take the values on the bosonic Matsubara frequencies. 
The key point in Eq.\ (\ref{eq:hui}) is analytic continuation in the chemical
potential. To perform it carefully, we proceed as follows.
From the definition of $\Z_{imp}$ and Eq.\ (\ref{Hamilton}) one  has:
\bee
{\partial\over\partial\lambda}\log\Z_{imp}
&=&
-T^{-1}\!\!\!\!\sum_{m=L,R}\langle
\psi^+_md+d^+\psi_m\rangle_{x=0} \label{eq:dod_Dyson} \\
&=&
-2\lambda\sum_{\omega_n}(\G_{0L}+\G_{0R})\G_{dd}^U(\omega_n),
\label{eq:another_one}
\eee
where $\G_{dd}^U=-\int_0^{1/T}d\tau e^{i\omega_n\tau}\langle 
T_{\tau}d(\tau)d^+(0)\rangle$ is the full Matsubara 
propagator for the impurity,
\be
\G_{dd}^U={1\over i\omega_n-\epsilon_0-\Gamma(\G_{0L}+\G_{0R})-\Sigma^M(
i\omega_n| \{i\mu\})},
\label{eq:trash}
\ee
$\omega_n$ are the fermionic Matsubara frequencies,
and to obtain (\ref{eq:another_one}) the Dyson equations were used,
which follow from the quadratic nature of coupling between the impurity
and  the leads:
\begin{mathletters}
\label{eq:all_Dys}
\bee
\langle T_{\tau}\psi_Ld^+\rangle_{x=0} &=& -\lambda\, \G_{0L}\, \G^U_{dd}, \\
\langle T_{\tau}d\psi_L^+\rangle_{x=0} &=& -\lambda\, \G_{0R}\, \G^U_{dd},
\eee
\end{mathletters}
where 
\be
\G_{0L(R)}=-\int_0^{1/T}d\tau e^{i\omega_n\tau}
\langle T_{\tau}\psi_{L(R)}(\tau)\psi_{L(R)}^+(0)\rangle_0
\label{DLDKDlla}
\ee
are free Matrubara Green functions
of the disconnected leads ($\lambda=0$) at $x=0$.
Substituting \cite{rem33}
\bee
\G_{0L(R)}(\omega_n)&=&\lim_{\Lambda\to\infty}
\int_{-\Lambda}^\Lambda {dp\over 2\pi}\, {1\over i\omega_n+p-i\mu_{L(R)}}
\nonumber \\
&=&-{i\over 2}\text{sign}(\omega_n-\mu_{L(R)}),
\label{eq:MatsI}
\eee
into (\ref{eq:another_one}) one has:
\bee
{\partial\over\partial\Gamma}\log\Z_{imp}=
i\sum_{\omega_n>0}&&\left[\G^U_{dd}(i\omega_n+i\max[\mu_L,\mu_R]) 
\right. \nonumber \\
&-& \left.
\G^U_{dd}(-i\omega_n+i\min[\mu_L,\mu_R]) \right],
\label{eq:comp_hui}
\eee
where it was used that $i\mu_{L(R)}$ are restricted to the 
  bosonic Matsubara frequencies, $2\pi iNT$, which allows to bring it in the
argument of the Green functions. Emergence of the non-analytic
functions ``max'' and ``min'' at this stage
of calculation is an important feature.
Expression (\ref{eq:comp_hui}) can be viewed as a function of two complex
variables $\mu_L$ and $\mu_R$ with a cut at $\mu_L=\mu_R$
\cite{rem76}. Because
of the cut, one can define two separate analytic continuations: once from
the domain $\mu_L>\mu_R$, and once from  $\mu_R>\mu_L$ to get two
different functions, related to $\Z_{imp}^+$ and $\Z_{imp}^-$ (this procedure is
analogous to obtaining retarded and advanced Green functions from the 
Matsubara Green function by continuation in $\omega$). Subtracting one 
function from another and substituting $i\mu_{L(R)}\to\mu_{L(R)}$, one obtains:
\bee
{\partial\over\partial\Gamma}\log{\Z_{imp}^+\over\Z_{imp}^-}
&=&
i\sum_{\omega_n>0}\left[\G^U_{dd}(i\omega_n+\mu_L)-\G^U_{dd}(i\omega_n+\mu_R)
\right. \nonumber \\
&+& \left.
\G^U_{dd}(-i\omega_n+\mu_L)-\G^U_{dd}(-i\omega_n+\mu_R)\right],
\label{eq:stam}
\eee
where the continuation $i\mu\to\mu$ is implicitly 
assumed in (\ref{eq:stam}) also 
in the second argument,
$\{\mu\}$, which is omitted here for brievity,
and the expression in the square brackets contains now only analytic
functions over real $\mu$.
 Note that
the sum over $\omega_n$ in (\ref{eq:comp_hui}) diverges
and hence an analytic continuation in $\mu$ might be not unique.
We assume the sum in (\ref{eq:comp_hui})
to be regularized in the standard way by inserting $\exp(i\omega_n\tau)$
\cite{AGD}. Then, since 
expression in the right-hand side of (\ref{eq:comp_hui})
is now a function of $\mu$ which vanishes when $|\mu|\to\infty$
(by the  reason similar to as a Green function 
vanishes when $i\omega_n\to\infty$),
the analytic continuation in $\mu$ to the appropriate domain
in the complex $\mu$ plane becomes unique by the well-known mathematical
theorem. One may also notice that, without the preliminary regularization, the
divergence in $\Z^+$ 
cancels anainst the same divergence in $\Z^-$
and the sum in (\ref{eq:stam}) is convergent, which also defines
a certain way of regularization. The correctness of this procedure
can be checked against the non-interacting case.

Next, consider an interacting non-equilibrium system in the real time.  
By definition,
\be
\hat{J}_L={\partial\rho_L\over\partial t}={i\over\hbar}[H,\rho_L],
\label{eqq:J}
\ee
where $\rho_L=e\psi^+_L\psi_L$ is the density of charge in the left lead. 
Opening 
the commutator in (\ref{eqq:J})  and taking the average, one obtains
\be
J_L={ie\lambda\over\hbar}\left(\langle d^+(t)\psi_L(t,0)\rangle - 
\langle\psi^+_L(t,0)d(t)\rangle\right).
\label{eqq:ZBHK}
\ee
Using Dyson equations (Eq.\ (\ref{my_eq}) from appendix B), in analogy
with the Matsubara calculation above,
correlators in (\ref{eqq:ZBHK}) can be expressed in terms of the free lead
propagators and the full $d$-level retarded Green function 
 to obtain  \cite{Nozier,Meir}:
\be
J=\case{1}{2}(J_L-J_R)={e\Gamma\over h}\Intg\!\! d\omega[f_L-f_R]
\text{Im}G_{dd}^R(\omega),
\label{eq:IJ}
\ee
where $f_{L,R}=f(\omega-\mu_{L,R})$ are the Fermi functions.
Eq.\ (\ref{eq:IJ})  was derived
without any reference to the nature of impurity interactions $H_{int}$; in the case
when $d$ has additional quantum numbers, trace operation must be performed
\cite{Meir}.
Full $d-d$ propagator,
\be
G^R_{dd}={1\over\omega-\epsilon_0+i\Gamma-\Sigma^R(\omega,\mu)},
\label{eq:full_prop}
\ee
can be obtained by the Keldysh technique \cite{LL} if one uses 
(\ref{eq:full_prop}) with $\Sigma^R=0$
as a zeroth-order approximation for the impurity Green function
and then adiabatically switches on interactions. Using that
$2i\, \text{Im}G^R=G^R-G^A$, and that $G^R\, (G^A)$ is analytic in
the upper (lower) complex half-plane, we split the expression for current
(\ref{eq:IJ}) into two parts, one containing $G^R$ and another $G^A$
and close the contour in the upper/lower complex half-planes respectively.
Then, by the residue theorem, one has:
\bee
J&=&{e\over h}\pi T\Gamma\sum_{\omega_n>0}[G_{dd}^R(\mu_R+i\omega_n)
- G_{dd}^R(\mu_L+i\omega_n) \nonumber \\ 
&+&
G_{dd}^A(\mu_R-i\omega_n)-G_{dd}^A(\mu_L-i\omega_n)]
\label{eq:sof-sof}
\eee
To this formula is related our first observation:
finding the exact quantum current is equivalent
 to evaluating the non-equlibrium
retarded Green function in the complex $\omega$ plane
at the fermionic Matsubara frequencies shifted
from the imaginary axis by the amount of the chemical potential.
Now it is convenient to compare the real time calculation with
the Matsubara claculation, Eq.\ (\ref{eq:stam}), since both
expressions are defined over the same set of points in the
complex $\omega$ plane $(\mu_{L(R)}+i\omega_n)$.
For the non-interacting case self-energy vanishes, 
so that one  has 
an exact expression
\be
G_{dd}^R={1\over \omega-\epsilon_0+i\Gamma}.
\label{eq:Retdd1}
\ee
Note that in the  absense of interactions the  retarded non-equilibrium
Green function does not depend on the  chemical potentials of  the leads
 and coincides
with the equilibrium retarded Green function. With interactions, the self-energy
(and  thus $G_{dd}^R$) appears to be in general bias-dependent. 
For the current one also has an exact result, given in Appendix A.
Comparing
(\ref{eq:stam}) with (\ref{eq:sof-sof}) for the free correlators 
$(\Sigma^M=\Sigma^R=0)$ we see that they agree {\it term by term}, hence
Eq.\ (\ref{eq:hui}) holds.
We checked also that the analogue of Eq.\ (\ref{eq:hui})
holds for other free models, e.g. the anisotropic two-channel Kondo model
in the Toulouse limit \cite{Appel,Hersh} or
 the point-contact model (see Appendix A):
\be
H_{pc}=H_0+\lambda\delta(x)(\psi_L^+\psi_R+\psi_R^+\psi_L).
\label{HPCa}
\ee
Hamiltonian (\ref{HPCa}) corresponds to a particular case of the quantum Hall bar
with an impurity \cite{FLS} at $\nu=1$, when the  Luttinger liquid on edges
reduces to free fermions, and thus it 
provides a direct check of Eq.\ (\ref{Sal}). 
However, in order to satisfy Eq.\ (\ref{eq:hui})
in the presence of interactions, one needs, e.\ g., 
\be
\Sigma^R(i\omega_n+\mu_L | \{\mu\})=\Sigma^M(i\omega_n+i\mu_L | \{i\mu\}
)_{i\mu\to\mu}
\label{eq:MAIN}
\ee
for $\omega_n>0$,
or, equivalently,
\be
\Sigma^R(i\omega_n | \{\mu\})=\Sigma^M(i\omega_n | \{i\mu\}_{i\mu\to\mu}).
\ee
 Equation (\ref{eq:MAIN}) states
that, out of equilibrium, if the retarded self-energy is
equal to the Matsubara self-energy analytically continued both
in $\omega$ and in $\mu$, then Eq.\ (\ref{eq:hui}) holds,
and it is very likely that for generic interactions  this statement works
both ways.
 Thus, Eq.\
(\ref{eq:MAIN}) represents a very interesting conjecture. It
is a very simple generalization of the analagous
result in the linear response theory. 
Indeed,
to obtain the current in the linear response limit,
$\mu_{L},\mu_R\to 0$
\cite{rem21}, one can substitute for the retarded Green function
in Eq.\ (\ref{eq:IJ}) its equilibrium value, and then   
Eq.\ (\ref{eq:MAIN}) becomes just a familiar relation from the linear response
theory. Hence, Eq.\ (\ref{eq:hui}) holds in the linear
response limit.

\section{Perturbative analysis out of equilibrium}

 In strongly interacting 1D systems a naive perturbation
series without some self-consistent resummations  often do not
reveal the correct physical properties of the system. However, such a series, if
convergent, are quite
suitable for the purposes of comparison of the mathematical structure
of both sides of (\ref{eq:MAIN}). 

 To see that in general out of equilibrium Eq.\ (\ref{eq:MAIN})  is not satisfied,
consider the simplest 
interacting model with $\{d_n\}=\{d_\uparrow, 
d_\downarrow\}$ and interaction of the form
\be
H_{int}=Ud^+_\uparrow d_\uparrow d^+_\downarrow d_\downarrow,
\label{eq:U}
\ee
(correspondingly, electrons in the leads also carry spin and interactions
 are such that the total spin is preserved). This is known as the
Anderson model \cite{ANDR}, and it was 
studied in the $U=\infty$ \cite{Meir2,Meir3,Hersh3} 
as well as small $U/\Gamma$
limits \cite{Hersh2} out of equilibrium for its connection with
quantum dots. In the equilibrium, Anderson model is exactly
solvable by the Bethe ansatz \cite{Wiegm},
and it is a potential candidate to be solved exactly out of  equilibrium.
 It can be shown \cite{ZH}
that, e.\ g., susceptibility in equilibrium can be expanded
into the power series in $U/\Gamma$ which converges absolutely
for $|U/\Gamma|<\infty$ and rapidly attains the asymptotic form. It is natural
to assume that such convergence properties remain true also out of equlibrium.
Thus, if Eq.\ (\ref{eq:MAIN}) holds, then it must be satisfied for every term 
in the $U$-perturbation expansion of both sides,
in analogy with the standard linear response theory \cite{YY}. We checked
Eq.\ (\ref{eq:MAIN}) for the first few terms. In the first order in $U$
the self-energy depends on the occupation of the levels at $U=0$.
The occupation $\langle n\rangle$
obtained     
from the analaytic continuation of the Matsubara Green function of
the non-interacting system, 
\bee
&&\Sigma_{(1)}^M=U\G^0(\tau,\tau+0)_{i\mu\to\mu}= \label{bbb} \\
&&{U\over 2} - {U\over\pi}\text{Im}\Psi\left({1\over 2}+i{\epsilon_0\over 2\pi T}
\right) + {U\over 2\pi i}\biggl[  \nonumber \\
&&\Psi\left({1\over 2}+{\Gamma\over 2\pi T}-i
{\epsilon_0+\mu\over 2\pi T}\right) -
\Psi\left({1\over 2}+{\Gamma\over 2\pi T}+i
{\epsilon_0-\mu\over 2\pi T}\right) \nonumber \\
&&+\Psi\left({1\over 2}+i{\epsilon_0-\mu\over 2\pi T}\right) -
\Psi\left({1\over 2}-i{\epsilon_0+\mu\over 2\pi T}\right) \biggr], \nonumber
\eee
 does not in general agree with the exact non-equilibrium
occupation of the same system, 
\bee
&&\Sigma^R_{(1)}=-iUG^{-+}(t,t)={U\over 2}-{U\over 2\pi}\text{Im}\biggl[
\label{eq:compare} \\
&&\Psi\left({1\over 2}+{\Gamma\over 2\pi T}+i{\epsilon_0+\mu\over 2\pi T}\right)
+\Psi\left({1\over 2}+{\Gamma\over 2\pi T}+i{\epsilon_0-\mu\over 2\pi T}\right)
\biggr], \nonumber
\eee
where $\Psi(x)$ is the derivative of logarithm of gamma-function and
we have taken $\mu=\mu_L=-\mu_R>0$.
We do not exclude, however, that for some special
choice of parameters Eq.\ (\ref{eq:MAIN}) might hold. Indeed, 
from the above explicit expressions in the first order we see that,
except for the equilibrium point $\mu=0$, the two expressions agree
also at the {\it symmetric point} $\epsilon_0=0$
out of equilibrium, where $\Sigma^R_{(1)}=
\Sigma^M_{(1)}=U/2$. Of course, one  cannot rely merely on the
first order result. It just indicates that  Eq.\ (\ref{eq:MAIN}) could hold
in the true symmetric point $\epsilon_0=-U/2$ where all the energy levels
$(\mu, -\mu, \epsilon_0, \epsilon_0+U)$ are symmetric with respect to zero
and an additional particle-hole symmetry is present. For $\epsilon_0=-U/2$
one treats the $\epsilon_0d^+d$ term as a perturbation, and 
the first order contribution to self-energy vanishes: $\Sigma^R_{(1)}=
\Sigma^M_{(1)}=0$, thus the non-trivial check for the symmetric point
can be provided by the second order in $U$. As can be shown,  
 only one diagram, Fig. 1, gives a non-vanishing contribution up to this order.
%%%%%%%%%%%%%%%%%%%%%%%%%%%%%%%%%%%%%%%%%%%%%%
\begin{figure}
\epsfxsize=25truemm
\centerline{\epsfbox{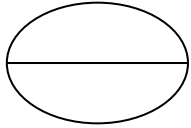}}
\vspace{5mm}
\caption{First non-vanishing contribution to the self-energy of
the symmetric  Anderson model.}
\end{figure}
%%%%%%%%%%%%%%%%%%%%%%%%%%%%%%%%%%%%%%%%%%%%%
Unfortunately, in the second order the calculation
is significantly more cumbersome and we were not able to 
complete it
and obtain an explicit analytical result. We hope, thus, that Eq,\ (\ref{eq:MAIN})
may hold for  the special choice of the value of $\epsilon_0$
in the Anderson model, leaving it as
an open question for future studies.
Note also
that according to the Luttinger-Ward procedure  for calculating
the ground state energy of interacting Fermi gas \cite{LW}
the first order correction
to the  self-energy (which is due to instantanious self-interaction) 
cancels against
the correction to chemical potential. This result does not apply to our case,
since the chemical potential is kept fixed in our problem, while the total
number of particles, being in fact infinite, is allowed to vary.

As another example, which allows to perform a reliable check
of Eq.\ (\ref{eq:MAIN})  avoiding above difficulties, we consider
model (\ref{Hamilton}) with the dissipation on impurity. Namely,
let $H_{int}$ describes the effects of phonons (photons) emission
and absorption on the impurity \cite{Frick},
\be
H_{int}=\gamma d^+d\phi,
\ee
where $\phi$ is a real phonon field
and $\gamma$ is the electron-phonon coupling constant. 
For  simplicity we consider a toy
model where the spectrum of phonons consists of a single mode,
$\omega_0(k)=\omega_0>0$. The phonons  are assumed to be in thermodynamic
equilibrium initially, with the energy
\be 
H_{ph}=\omega_0b^+b.
\ee
The field $\phi$ is quantized in terms of the bosonic operator $[b,b^+]=1$ as
follows:
\be
\phi=i\sqrt{{\omega_0\over 2}}(b-b^+).
\ee
The first non-trivial correction to the $d$-electron self-energy is
given by the one-loop diagrams shown in Fig.\ 2. 
%%%%%%%%%%%%%%%%%%%%%%%%%%%%%%%%%%%%%%%%%%%%
\begin{figure}
\epsfxsize=50truemm
\centerline{\epsfbox{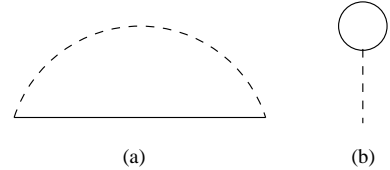}}
\vspace{5mm}
\caption{The leading contributions to the self-energy of d-electron
interacting with phonons (dashed line).}
\end{figure}
%%%%%%%%%%%%%%%%%%%%%%%%%%%%%%%%%%%%%%%%%%%%%
The diagram in Fig.\ 2(b)
is proportional to $\G^0_{dd}(\tau,\tau+0)$ and thus  reduces to the analysis
of the first order correction for the Anderson model (\ref{bbb}). The diagram
shown in Fig.\ 2(a), however,
contains a non-trivial $\omega$ and $\mu$ dependence 
and can be easily evaluated analytically.  
One has
\be
\Sigma^M_{(1)}(i\omega_n, i\mu)=-\gamma^2T\sum_{\Omega_p}
\D(\Omega_p)\G_{dd}^0(\omega_n+\Omega_p),
\label{eq:one}
\ee
where $\Omega_p$ is the bosonic Matsubara frequency and $\D(\tau)=
-\langle T_\tau\phi(\tau)\phi(0)\rangle$ is the phonon Matsubara propagator,
\be
\D(\Omega_p)=-{\omega_0^2\over\Omega_p^2+\omega_0^2}.
\ee
The corresponding non-equilibrium retarded self-energy (Fig.\ 3)
is given by
\bee
\Sigma^R_{(1)}(\omega,\mu)&=& i\gamma^2\Intg{d\omega_1\over 2\pi}
\bigl[D^{- -}(\omega_1)G^{- -}_{dd}(\omega_1+\omega) \nonumber \\
&-& D^{+-}(\omega_1)G^{-+}_{dd}(\omega_1+\omega)\bigr],
\label{eq:two}
\eee
where the Keldysh correlators $D^{- -}, D^{+-}$ are given in Appendix B.
Substituting corresponding bare propagators and evaluating
(\ref{eq:one}),(\ref{eq:two}) explicitly one  can see that the only case
when Eq.\ (\ref{eq:MAIN}) holds is the case of equilibrium, $\mu=0$.
%%%%%%%%%%%%%%%%%%%%%%%%%%%%%%%%%%%%%%%%%
\begin{figure}
\epsfxsize=60truemm
\centerline{\epsfbox{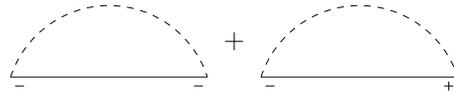}}
\vspace{5mm}
\caption{Leading contribution to the retarded self-energy of d-electron
interacting  with phonons, $\Sigma^R=\Sigma^{-+}+\Sigma^{- -}$,
non-equilibrium case.}
\end{figure}
%%%%%%%%%%%%%%%%%%%%%%%%%%%%%%%%%%%%%%%%%%%%%

\section{Conclusion}
\label{Concl}
To summurize, a generalization of the linear response theory
to the non-equilibrium case was discussed. 
This generalization is based on the certain analytic continuation
in the additional complex parameter which plays a role of external bias.
The complex chemical potential is formally introduced in the equilibrium density
matrix and analytically continued to the real values after averaging over
all the states is performed. Thermodynamic averages turn out to have
non-analytic dependence on the complex bias, with the cut along the plane
$\text{Im}\mu_L=\text{Im}\mu_R$, as a result of divergences in the summation
over infinite-particle states. 
It was shown that
formula for the current (\ref{Sal}) is valid in the linear
response limit for systems of the type
(\ref{Hamilton}) and is valid also out of equilibrium if
condition (\ref{eq:MAIN}) holds. This condition, which is trivially
satisfied by the free systems, $H_{int}=0$, is not satisfied in general
by the interacting systems. This does not
mean, however, that the very idea of obtaining non-equillibrium current
by exploiting analytic properties of the thermodynamic averages in the 
space of complex coupling constants is hopeless. The fact that Eq.\ (\ref{Sal})
holds for the free systems indicates that there are no general grounds
for ruling out this idea. 
As the perturbative analysis indicates, the propagators obtained
by an analytic continuation from the Matsubara ones have similar structure
and resemble real non-equilibrium propagators.
Hence, it would be interesting to derive an 
analogue of Eq.\ (\ref{Sal}) based
on the analytic continuation in the complex bias
which would be true for the generic interacting
systems as well. One may proceed, e.\ g., from the Lehmann spectral
representation \cite{Zub} to obtain a relation between the 
equilibrium partition function and the non-equilibrium current
by controlling two complex parameters, frequency and chemical potential.
Formula (\ref{Sal}) as it is can be used, nevertheless, to obtain the exact transport
through Anderson impurity in the linear response limit \cite{GR}, if the
exact partition function is supplied by the Bethe ansatz technique. 
It would be interesting to understand also 
in the Green functions language why Eq.\ (\ref{Sal}) works for the quantum
Hall bar with a constriction \cite{FLS}.

%%%%%%%%%%%%%%%%%%%%%%%%%%%%%%%%%%%%%%%%%%%%%%%%%%%%%%%%%%%%%%%%%%%
\acknowledgments
%%%%%%%%%%%%%%%%

Discussions with Natan Andrei, Andrei Dashevsky,
Alexander Finkelstein, Baruch Horowitz and Yigal Meir were very helpful
and greatly contributed to this work.

The author was supported by Kreitman and by Kaufmann fellowships
and also partially funded by the Israel Science Foundation, grant
No 3/96-1.

%%%%%%%%%%%%%%%%%%%%%%%%%%%%%%%%%%%%%%%%%%%%%%%%%%%%%%%%%%%%%%%%%%%%%%%%

\appendix
%%%%%%%%%%
\section{Free case}
\label{sec:free}
\subsection{Point-contact model}

To understand how the analytic continuation in
the complex chemical potential works, we consider the following
simple  point-contact model \cite{rem2}:
\be
H_{pc}=H_0+\lambda\delta(x)(\psi_L^+\psi_R+\psi_R^+\psi_L).
\label{HPC}
\ee
This model can be viewed also as a particular case of the quantum Hall bar
with an impurity \cite{FLS} at $\nu=1$, when the  Luttinger liquid on edges
reduces to free fermions, and thus it 
provides a direct check of Eq.\ (\ref{Sal}).
By definition,
\be
\hat{J}_L={\partial\rho_L\over\partial t}={i\over\hbar}[H_{pc},\rho_L],
\label{eq:J}
\ee
where $\rho_L=e\psi^+_L\psi_L$ is the density of charge. Opening the commutator
above and taking the average, one obtains
\be
J_L={ie\lambda\over\hbar}\left(\langle\psi^+_R\psi_L\rangle-
 \langle\psi^+_L\psi_R\rangle\right).
\label{eq:JJ}
\ee
The latter expression involves equal time correlators at $x=0$,
which can be easily found by using Dyson equations and the Keldysh technique
\cite{LL}. Eventually, one derives
\bee
J_L&=&{e\lambda^2\over h}\int_{-\infty}^\infty\!\! d\omega[f_L-
f_R]{2\text{Im}g^R\over|1-(\lambda g^R)^2|^2} 
 \label{eq:JJJ} \\
&=&{e\over h}{\Gamma\over(1+\Gamma/4)^2}\left(\mu_R-\mu_L\right)
\label{eq:JJJJ}
\eee
where $f_{L,R}=f(\omega-\mu_{L,R})$ are the Fermi functions, 
$\Gamma=\lambda^2$ and
$g^R$ is the retarded Green function of either of the disconnected leads
$(\lambda=0)$ at $x=0$ \cite{rem3}, 
\be
g^R(\omega)=\Intg {dp\over 2\pi}\,{1\over \omega+p+i0}=-{i\over 2}.
\label{eq:Ret}
\ee

Partition function of (\ref{HPC}) can be easily found.  
For the complex chemical potentials $i\mu_L$ and $i\mu_R$, 
according to (\ref{DefPF}) one has
\bee
{\partial\over\partial\Gamma}&&\!\!\!\!\!\!\!
\log\Z_{imp}(i\mu_{L},i\mu_R)=-\!\!\!\sum_{\omega_n}
{\G_{0L}(\omega_n)\G_{0R}(\omega_n)\over 1-\Gamma
\G_{0L}(\omega_n)\G_{0R}(\omega_n)} \nonumber \\
&=&{|\mu_L-\mu_R|\over 2\pi T}{1\over\Gamma-4}+
\sum_{\omega_n>\max\{\mu_L,\mu_R\}\atop\omega_n<\min\{\mu_L,\mu_R\}}
{1\over\Gamma+4},
\label{eq:ASSA}
\eee
where 
\be
\G_{0L(R)}=-\int_0^{1/T}d\tau e^{i\omega_n\tau}
\langle T_{\tau}\psi_{L(R)}(\tau)\psi_{L(R)}^+(0)\rangle_0
\label{DLDKDll}
\ee
 are the Matsubara Green functions for the disconnected leads
$(\lambda=0)$ at $x=0$, e.\ g. 
\bee
\G_{0L}(\omega_n)&=&
\Intg {dp\over 2\pi}\, {1\over i\omega_n+p-i\mu_L}
\nonumber \\
&=&-{i\over 2}\text{sign}(\omega_n-\mu_{L}),
\label{eq:Mats}
\eee
and $\omega_n$ are the fermionic Matsubara frequencies. When passing from
the first line to the second in (\ref{eq:ASSA}) it was used that 
 $\mu_L=2\pi NT$ and $\mu_R=2\pi MT$, where $N,M$ are integers. 
The infinite sum in (\ref{eq:ASSA})  needs to be properly regularized:
\bee
\sum_{\omega_n>\max[\mu_L,\mu_R]\atop\omega_n<\min[\mu_L,\mu_R]}\!\!\!
{1\over\Gamma+4}&=&-{1\over 2\pi iT}{1\over\Gamma+4}\lim_{\epsilon\to+0}
\oint_C{e^{\epsilon\omega}\over e^{\omega/T}+1}d\omega \nonumber \\
&=& {|\mu_L-\mu_R|\over 2\pi T}{1\over\Gamma+4},
\label{eq:SLEEP}
\eee
where contour $C$ consists of two parts: one goes along $\text{Im}\, \omega
=\max[\mu_L,\mu_R]$ and an infinite semicircle above it, the other
along $\text{Im}\, \omega=\min[\mu_L,\mu_R]$ and a semicircle below it,
both counter-clockwise. 

Considering $\mu_L-\mu_R$ as one variable $y$, one has a function
$|y|$ defined on the imaginary $y$-axis. We have to continue it analytically
to the real axis. Since it has a cut at $y=0$, one can define two
separate analytic continuations, to the upper and lower half-planes:
$f_+=-iz$, $f_-=iz$. Then, one continues from the positive imaginary axis
to the real axis, $f_+(x)=-ix$, and from the negative imaginary axis,
$f_-(x)=ix$, and subtracts one from another to get $f_+-f_-=-2i(\mu_L-\mu_R)$
\cite{rem4}.
Comparing to (\ref{eq:JJJJ}), one has
\be
J={e\over h}8\pi iT{4-\Gamma\over 4+\Gamma}{\partial\over\partial\Gamma}\log
{\Z_{imp}^+\over\Z_{imp}^-}.
\label{eq:endresI}
\ee
Note that the prefactor $(4-\Gamma)/(4+\Gamma)$ in (\ref{eq:endresI})
is not universal and depends on the regularization scheme.

\subsection{Resonant-level model}

Next, instead of the point-contact scattering we introduce
a single dynamical impurity at $x=0$ described by an additional state at the
energy $\epsilon_0$. Operator $d^+$ creates an electron on the impurity.
The simplest model to write is a resonant-level model:  
\be
H_{RL}=H_0+\epsilon_0d^+d+\lambda\!\!\!\!\!
\sum_{m=L,R}(\psi^+_m(0)d+d^+\psi_m(0)).
\label{HRL}
\ee
Note that 
the impurity can be effectively removed from the action by integrating
out impurity degrees of freedom, and one gets certain generalization
of the point-contact model with time-dependent coupling. Then, in the
limit $\epsilon_0,\lambda\to\infty$ one recovers (\ref{HPC}).
A dynamical impurity, however,  suggests a solid way of 
regularizing the divergences
encountered in Eq.\ (\ref{eq:ASSA}). 

Expression for the current was derived in Sec.\ \ref{sec:inter}:
\be
J=\case{1}{2}(J_L-J_R)={e\Gamma\over h}\Intg\!\! d\omega[f_L-f_R]
\text{Im}G_{dd}^R(\omega),
\label{eq:IJk}
\ee
where  for the retarded Green
function $ G_{dd}^R$ one has in this case an exact expression
\be
G_{dd}^R={1\over \omega-\epsilon_0+i\Gamma}.
\label{eq:Retdd}
\ee
Substituting (\ref{eq:Retdd}) into (\ref{eq:IJk}) one obtains:
\bee
J&=&{e\over h}\Gamma\,
\text{Im}\biggl[\Psi\left({1\over 2}+{\Gamma\over 2\pi T}
+i{\epsilon_0-\mu_L\over 2\pi T}\right)  \nonumber \\
&-&\Psi\left({1\over 2}+{\Gamma\over 2\pi T}
+i{\epsilon_0-\mu_R\over 2\pi T}\right)\biggr],
\label{eq:DIGAMMA}
\eee
where $\Psi$ is the standard psi (digamma) function (compare to 
Eq.\ (\ref{eq:compare})). 

The partition function involves gaussian Feynmann integral
and can be obtained immediately:
\bee
{\partial\over\partial\Gamma}\log&\Z_{imp}&(i\mu)=
-\sum_{\omega_n}{\G_{0L}(\omega_n)+\G_{0R}(\omega_n)\over
i\omega_n-\epsilon_0-\Gamma(\G_{0L}+\G_{0R})} \nonumber \\
= \sum_{\omega_n>0}&&\biggl[
{i\over i\omega_n+i\max[\mu_L,\mu_R]-\epsilon_0+i\Gamma} 
\nonumber \\
&+&{i\over i\omega_n-i\min[\mu_L,\mu_R]+\epsilon_0+i\Gamma}\biggr]
\label{eq:PFunck}
\eee
where, while passing from the first line to the second above,
 Eq.\ (\ref{eq:Mats}) was used and the fact that $i\mu_{L,R}$ have the
form of bosonic Matsubara frequencies, which allows to bring $i\mu_{L,R}$
into the denominators. Expression (\ref{eq:PFunck}), viewed as a function
of two complex variables $\mu_L$ and $\mu_R$, has a cut at $\mu_L=\mu_R$.
The analytic continuation is performed from the 2D plane of purely
imaginary $\mu_L$ and $\mu_R$ with the cut at  $\mu_L=\mu_R$. Once
one continues analytically from the domain $\mu_L>\mu_R$, and
once from $\mu_R>\mu_L$ to get two functions. Subtracting
one from another and substituting $i\mu\to\mu$, one gets by comparing
with (\ref{eq:DIGAMMA}) 
\be
J={e\over h}i\pi T\Gamma{\partial\over\partial\Gamma}\log{
\Z_{imp}^+\over\Z_{imp}^-}.
\label{eq:huija}
\ee

\subsection{Kondo model}
\label{sec:Kondo}

Consider an impurity spin $S=\case{1}{2}$
which couples $L$ and $R$ leads by the $s-d$ exchange (Kondo) interaction
\cite{Appel}:
\be
H_K=H_0+\sum_{\lambda=1}^3\sum_{ab=L,R}J^{ab}_\lambda\psi^+_{\sigma a}(0)
\sigma^\lambda_{\sigma\sigma'}\psi_{\sigma' b}(0)S^\lambda,
\label{HK}
\ee
where $\sigma^\lambda$ are Pauli matrices and $J^{ab}_\lambda$ are
coupling constants. Electrons here have additional spin index $\sigma$.
As it was shown in \cite{Hersh} following the ideas of \cite{EK},
under the assumptions that $J_z^{LR}=J_z^{RL}=0,\: J_x=J_y=J_\bot, \: 
J_\bot^{LL}=J_\bot^{RR}, \: J_\bot^{LR}=J_\bot^{RL}, \:
    J_z^{LL}=J_z^{RR}=2\pi$ model (\ref{HK})
is equivalent  to a quadratic model with the Hamiltonian
\bee 
H_K&=&\sum_{m=1,2}i\!\!\int dx\psi^+_m\partial_x\psi_m
+J_1\delta(x)(\psi^+_1+\psi_1)(d^+-d) \nonumber \\
&+&J_2\delta(x)(\psi^+_2-\psi_2)(d^++d),
\label{eq:mudak}
\eee
where $J_1\sim J_\bot^{LL}$,  $J_2\sim J_\bot^{LR}$, and $\{d^+,d\}=0$.
Indexes 1 and 2 correspond to the spin-flavor and flavor
channels of the original model (\ref{HK}) \cite{EK}, while only
second channel has now non-zero chemical potential $\mu$ if one
sets initially $\mu_L=-\mu_R=\case{\mu}{2}$. When calculating
$\Z_{imp}$ it is convenient to change variables in the Fenmann integral
from $d,d^+$ to the real and imaginary parts $a=(d+d^+)/\sqrt{2}, \;
b=i(d-d^+)/\sqrt{2}$, where $a$ and $b$ are classical Majorana fermions.
After integrating out $\psi_{1,2}$ the action factorizes as
$S=S(a,J_2^2,\mu)+S(b,J_1^2)$, and only the $\mu$-dependent part is
of interest to us. Since $S$ is gaussian, $\Z_{imp}$ can be easily computed.
The result coincides with Eq.\ (\ref{eq:PFunck}) at $\epsilon_0=0,\:  \mu_L=
-\mu_R$. Thus, Kondo interactions reduce to the case of resonant-level
model studied above for the special point in the couplings space
(an analogue of Toulouse point for the single-channel Kondo problem).

\section{Green functions}

We  collect here the definitions of different Green functions
used throughout this Letter, following the notations of Ref.\ \cite{LL}.
All the Green functions are taken at $x=0$. The subscript $R$
for ``right'' should not be confused with superscript $R$ for ``retarded''.
 First, consider the point-contact model.
Define:
\bee
i\langle\psi^+_R(t)\psi_L(t)\rangle&=&G^{-+}_{RL}(0)
\\
i\langle\psi^+_L(t)\psi_R(t)\rangle&=&G^{-+}_{LR}(0)
\eee
Dyson equations yield:
\begin{mathletters}
\bee
G^{-+}_{LR}&=&\lambda g^{- -}_R G_{LL}^{-+} - \lambda g^{-+}_R G_{LL}^{++}
\\
G^{-+}_{RL}&=&\lambda g^{- -}_L G_{RR}^{-+} - \lambda g^{-+}_L G_{RR}^{++}
\eee
\end{mathletters}
where
\bee
-ig^{-+}_{L(R)}(\omega)&=&\Intg \!\!\! dt e^{i\omega t}
\langle\psi^+_{L(R)}(0)\psi_{L(R)}(t)\rangle \nonumber \\ 
&=& f(\omega-\mu_{L(R)}), 
\eee
\bee
ig^{+-}_{L(R)}(\omega)&=&\Intg \!\!\! dt e^{i\omega t}
\langle\psi_{L(R)}(t)\psi^+_{L(R)(0)}\rangle \nonumber \\
&=& [1-f(\omega-\mu_{L(R)})] 
\eee
\be
ig^{- -}_{L(R)}(\omega)=\Intg \!\!\! dt e^{i\omega t}
\langle T_t\psi_{L(R)}(t)\psi^+_{L(R)}(0)\rangle
\ee
\be
ig^{++}_{L(R)}(\omega)=\Intg \!\!\! dt e^{i\omega t}
\langle \tilde{T}_t\psi_{L(R)}(t)\psi^+_{L(R)}(0)\rangle
\ee
\be
g^R=\overline{g^A}=-{i\over 2}
\ee
are Green functions of free (decoupled, $\lambda=0$) leads,
$\tilde{T}_t$ stands for  anti time ordering operation,
and by capital letters $G_{LL}, G_{RR}$ are denoted corresponding
full propagators of the interacting system ($\lambda\neq 0$).
The latter can be found from the system of four coupled Dyson equations.
The solution reads, e.\ g.,
\bee
G_{LL}^{-+}&=&
{g_L^{-+}+\lambda^2 g_L^R g_L^A g_R^{-+} \over
|1-\lambda^2 g_L^R g_R^R|^2}
\\
G_{LL}^{++}&=&
{g_L^{++}+\lambda^2 g_L^R g_L^A g_R^{- -} \over
|1-\lambda^2 g_L^R g_R^R|^2}
\eee
For the resonant-level model there are very similar definitions.
For example,
\bee i\langle d^+(t)\psi_L(t)\rangle &=& G^{-+}_{dL}(0) \\
  i\langle\psi^+_L(t) d(t)\rangle &=& G^{-+}_{Ld}(0)\eee
Dyson equations express these correlators through
the full propagator of the impurity:
\begin{mathletters}
\label{my_eq}
\bee
G^{-+}_{dL}&=&\lambda g_L^{- -} G_{dd}^{-+} - \lambda g_L^{-+} G_{dd}^{++}
\\
G^{-+}_{Ld}&=&\lambda G_{dd}^{- -} g_L^{-+} - \lambda G_{dd}^{-+} g_L^{++}
\eee
\end{mathletters}
Four Green functions $++, - -, +-, -+$ in general are related to
each other by
\be
G^{++}+G^{- -}=G^{+-}+G^{-+}
\ee
and to the retarded (advanced) Green functions
\bee
G^R=G^{- -}-G^{-+}, \\
G^A=G^{- -}-G^{+-}.
\eee
The  impurity Green functions of an isolated non-interacting impurity read:
\be
G_{dd}^{-+(0)}=i\pi\delta(\omega-\epsilon_0)
\ee
\be
G_{dd}^{+-(0)}=-i\pi\delta(\omega-\epsilon_0)
\ee
\be
G^{R(0)}_{dd}={1\over \omega-\epsilon_0 +i0}
\ee
For the resonant-level model without on-site interactions 
one can write four Dyson equations for the impurity
correlators and easily solve them.
Relevant for us is the retarded Green function, 
\be
G_{dd}^R= {G_{dd}^{R(0)}\over 1-G_{dd}^{R(0)}\Sigma^{R}}
\ee
where $\Sigma^R=\Sigma^{- -}+\Sigma^{-+}$ is the retarded self-energy,
given in the free case by
\be
\Sigma^{R(0)}=\lambda^2 (g_L^R+g_R^R)
\ee
For the calculation of  interacting Green functions one needs
also the following zeroth-order propagators:
\be
G_{dd}^{-+}={G_{dd}^{-+(0)}+\lambda^2 G^{R(0)}_{dd}G^{A(0)}_{dd}
(g_L^{-+}+g_R^{-+})\over |1-\lambda^2 G^{R(0)}_{dd}(g_L^{R}+g_R^{R})|^2}
\ee
\be
G_{dd}^{+-}={G_{dd}^{+-(0)}+\lambda^2 G^{R(0)}_{dd}G^{A(0)}_{dd}
(g_L^{+-}+g_R^{+-})\over |1-\lambda^2 G^{R(0)}_{dd}(g_L^{R}+g_R^{R})|^2}
\ee
\be
G_{dd}^{- -}={G_{dd}^{- -(0)}+\lambda^2 G^{R(0)}_{dd}G^{A(0)}_{dd}
(g_L^{++}+g_R^{++})\over |1-\lambda^2 G^{R(0)}_{dd}(g_L^{R}+g_R^{R})|^2}
\ee
Zeroth order correlators of the phonon field are defined as
\bee
iD^{-+}(t_1,t_2)&=&\langle\phi(t_2)\phi(t_1)\rangle \\
iD^{+-}(t_1,t_2)&=&\langle\phi(t_1)\phi(t_2)\rangle
\eee 
Explicitly,
\bee
D^{-+}(\omega)&=&-i\pi\omega_0[N\delta(\omega-\omega_0)+
(1+N)\delta(\omega+\omega_0)] \\
D^{+-}(\omega)&=&-i\pi\omega_0[N\delta(\omega+\omega_0)+(1+N)
\delta(\omega-\omega_0)] \\
D^R(\omega)&=&{\omega_0\over 2}\left[{1\over \omega - \omega_0 +i0} -
 {1\over \omega + \omega_0 +i0}\right]
\eee 
with
\be 
N={1\over e^{\omega_0/T}-1}
\ee
%%%%%%%%%%%%%%%%%%%%%%%%%%%%%%%%%%%%%%%%%%%%%%%%%%%%%%%%%%%%%%%%%%%%%%

%\pra\prb\prl\rmp\pl\jpp - shorthands for journals 


\begin{references}
\bibitem{conj} Since it is not a rigourous proof, formula  (\ref{Sal}) 
appeared as a well-checked conjecture.
\bibitem{rem1} Final result for the current does not depend on the choice
of branches.
\bibitem{rem22} Here $\Gamma$ has the units of energy, $\lambda$ being
the renormalized
coupling constant of the field theory (\ref{Hamilton}). Usually one starts
with the system in the box and the hopping  $\lambda_1\sum_{k}
c^+_kd$. Then $\lambda\sim D^{-1/2}\lambda_1$ where $D$ is a bandwidth.
\bibitem{rem33} 
All the expressions in this paper are given in the form when the
cutoff has been removed  already and  the issues related to the regularization
are not discussed. Nevertheless, it would  be interesting  to analyze
different regularization schemes and repeat the  calculation with a finite
analytical particle-number preserving cutoff.
\bibitem{rem76} Without loss of generality, one can set $\mu=\mu_L, \mu_R=0$
and deal with a function of one complex variable $\mu$, with the cut 
along the real axis.
\bibitem{rem21} Since  the current depends on the difference $\epsilon_0-\mu$,
by the overall energy shift one can always shift
$\mu$ to zero in equilibrium, measuring all the energies from
$\mu=0$. The relevant adjustable parameter is then  $\epsilon_0$.
\bibitem{rem2} Here $\lambda$ is the dimensionless renormilized coupling
constant of the field  theory (\ref{HPC}). Usually one starts with the system
in the box and the interaction $\lambda_1\sum_{k,k'}c_k^+c_{k'}$.
Then $\lambda\sim D^{-1}\lambda_1$ where $D$ is a bandwidth.
\bibitem{rem3} Integral in (\ref{eq:Ret}) can be regularized either
by introducing finite bandwidth cutoff, or by infinitesimal
splitting between the spatial arguments of correlator,  $\cos(px)$.
\bibitem{rem4} $f_+$ and $f_-$ are analogous to retarded and advanced Green
functions.
\end{references}
\end{document}